\documentclass[11pt]{article}

\usepackage{breqn}
\usepackage{color}
\usepackage{bm}
\usepackage{multirow}
\usepackage[ruled,linesnumbered]{algorithm2e}
\usepackage{graphicx}

\newcommand{\norm}[1]{\left\lVert#1\right\rVert}
\newcommand{\fig}[4]{
    \begin{figure}[tb]\centering
    \includegraphics[width=#4in]{#1}
    \caption{#2}
    \label{#3}\end{figure}
    {}}

\usepackage{subfigure}
\newcommand{\figtwo}[8]{
    \begin{figure}[tb]\centering
    \subfigure[#2] {
    \label{#3}
    \includegraphics[width=2.3in]{#1}
    }
    \subfigure[#5] {
    \label{#6}
    \includegraphics[width=2.3in]{#4}
    }
    \caption{#7}
    \label{#8}
    \end{figure}
    {}}

\begin{document}

\title{Weighted-SVD: Matrix Factorization with Weights on the Latent Factors}

\author{Hung-Hsuan Chen \\
       hhchen@ncu.edu.tw \\
       Department of Computer Science and Information Engineering\\
       National Central University\\
       }

\maketitle

\begin{abstract}%   <- trailing '%' for backward compatibility of .sty file

The Matrix Factorization models, sometimes called the latent factor models, are
a family of methods in the recommender system research area to (1) generate the
latent factors for the users and the items and (2) predict users' ratings on
items based on their latent factors.  However, current Matrix Factorization
models presume that all the latent factors are equally weighted, which may not
always be a reasonable assumption in practice.  In this paper, we propose a new
model, called Weighted-SVD, to integrate the linear regression model with the
SVD model such that each latent factor accompanies with a corresponding weight
parameter.  This mechanism allows the latent factors have different weights to
influence the final ratings.  The complexity of the Weighted-SVD model is
slightly larger than the SVD model but much smaller than the SVD++ model.  We
compared the Weighted-SVD model with several latent factor models on five
public datasets based on the Root-Mean-Squared-Errors (RMSEs).  The results
show that the Weighted-SVD model outperforms the baseline methods in all the
experimental datasets under almost all settings.

\end{abstract}

%\begin{keywords}
%  Matrix factorization, latent factor model, recommender systems, SVD, linear regression
%\end{keywords}

\section{Introduction}

Recommender systems are widely used in modern days~\cite{liu2010personalized,
huang2014refseer, chen2011collabseer, chen2015expertseer, tang2008arnetminer}.
Among various techniques in recommender systems, Matrix Factorization (MF),
also known as the latent factor model, is one popular family of methods to
predict users' preferences on items based on users' historical actions on the
items.  Compared to the traditional collaborative filtering approaches, which
typically based on the concept of $k$-nearest-neighbors, MF discovers the
vector of the latent factors for each user $i$ and each item $j$ and assumes
that their interaction (i.e., the inner-dot operation) influences $i$'s final
rating on $j$.  Additionally, MF can incorporate other information, such as
user $i$'s bias and item $j$'s bias, the implicit feedback of $i$ on $j$, and
the temporal dynamics, to improve the quality of the
predictions~\cite{koren2008factorization, koren2009matrix,
mnih2008probabilistic}.  Moreover, we may apply various optimization
techniques, such as stochastic gradient descent (SGD), AdaGrad, Adam, or Nadam,
to obtain the latent factors and other parameters.  As a result, MF and its
extensions are efficient, effective, easy to implement, and domain-independent.

Despite of the various advantages, we found that MF-based approaches have one
fundamental assumption that might be too na\"{i}ve -- each latent factor has
equal influence power to the final rating.  Although we cannot ensure the
essential meaning of each latent factor, it seems more possible that some
latent factors have larger impacts to users' preferences on items compared to
the other latent factors.  This motivates us to develop the Weighted-SVD (WSVD)
model, which integrates the linear regression model with the SVD model, a
popular method in the family of the latent factor models.  The WSVD model
assigns a weight parameter to each latent factor.  During the training process,
WSVD learns not only the latent factors for users and items but also the
corresponding weights of the hidden factors.  We conducted extensive comparison
on the WSVD model with several latent factor models as baselines -- the SVD
model, the SVD++ model, and the PMF  (Probabilistic Matrix Factorization)
model.  We found that the WSVD model outperforms all the baselines under all
the experimental datasets, including three MovieLens datasets (MovieLens-100K,
MovieLens-1M, MovienLens-10M), FilmTrust dataset, and the Epinions dataset.
Additionally, we compared various settings of the hyper-parameters on various
latent factor models based on the MovieLens-100K dataset.  The results
demonstrate that WSVD consistently beat the other methods under different
settings.

The rest of the paper is organized as follows.  In Section~\ref{sec:method}, we
show the formulation of the WSVD model, the parameter learning process, and the
complexity of the model.  Section~\ref{sec:exp} reports the experiments to
compare the prediction performance and the training time of WSVD and other
methods.  Section~\ref{sec:related-work} shows the related works on recommender
systems and latent factor models.  Finally, we discuss the limitations of WSVD
and future research directions in Section~\ref{sec:disc}.

\section{Methodology}\label{sec:method}

This section shows the WSVD formula, the learning process,  the algorithm of
the WSVD model, and analyze the complexity of the WSVD model.

\subsection{Preliminaries}

The simplest rating prediction method is probably the average model -- setting
each un-rated score as the average of all the known ratings.  Thus, the average
model set $\hat{r}_{ij}$ the predicted rating of user $i$ on item $j$ as
$\hat{r}_{ij} = \sum_{\forall (x,y) \in \mathcal{K}} r_{xy} / |\mathcal{K}|$,
in which $r_{ij}$ is the rating of user $i$ on item $j$, $\mathcal{K}$ is the
set of all known (i.e., rated) (user, item) pairs, and $|\mathcal{K}|$ is the
size of the set $\mathcal{K}$, i.e., the number of rated pairs.

The bias model enhances the average model by adding the user and the item
biases.  A user who tends to rate scores higher than average would have a
positive user bias.  Similarly, an item has a positive item bias if the
received ratings are usually higher than average.  Equation~\ref{eq:bias-model}
shows the formula of the bias model.

\begin{dmath} \label{eq:bias-model}
\hat{r}_{ij} = \bar{r} + b_i^{(U)} + b_j^{(I)},
\end{dmath}
in which $\bar{r}$ is the average of all the rated scores, $b_i^{(U)}$ is the
user bias of user $i$, and $b_j^{(I)}$ is the item bias of item $j$. Specifically,
$b_i^{(U)}$ is set as user $i$'s average ratings on items subtracts $\bar{r}$
the average scores of all known ratings, and $b_j^{(I)}$ is set as the average
ratings received by the item $j$ subtracts $\bar{r}$.  Thus, the bias model
includes each user's rating tendency (i.e., tend to over-rate or under-rate the
items) and each item's overall quality compared to other items (i.e., an item
tends to receive ratings higher or lower than the average).

The SVD model introduced in~\cite{koren2008factorization} further improves the
bias model by introducing the inner-product term of user's and item's latent
factors.  Specifically, a user $i$ is associated with $\boldsymbol{p}_i$ a
vector of user $i$'s latent factors, and an item $j$ is associated with
$\boldsymbol{q}_j$ a vector of item $j$'s latent factors.  Both
$\boldsymbol{p}_i$ and $\boldsymbol{q}_j$ are column vectors of size $k$, which
represents the number of the latent factors and needs to be specified
beforehand.  The model presumes that the product of each latent factor in
$\boldsymbol{p}_i$ and the corresponding latent factor in $\boldsymbol{q}_j$
also affects user $i$'s ratings on item $j$, as shown in
Equation~\ref{eq:latent-model}.

\begin{dmath} \label{eq:latent-model}
\hat{r}_{ij} = \bar{r} + b_i^{(U)} + b_j^{(I)} + \boldsymbol{p}_i^T \cdot \boldsymbol{q}_j.
\end{dmath}

%Note that, different from the bias model, in the SVD model the user bias
%$b_i^{(U)}$ and the item bias $b_j^{(I)}$ are also unknown paramters need to be
%learned.

The SVD model can further be improved by including the implicit feedback,
users' features, temporal dynamics, etc.  Many of these techniques are
discussed in~\cite{koren2009matrix}.  Here we introduce the SVD++ model -- a
model incorporates the implicit feedback that indirectly reveals users'
opinions on the items.  Equation~\ref{eq:svd++-model} shows the equation to
predict user $i$'s rating on item $j$.

\begin{dmath} \label{eq:svd++-model}
\hat{r}_{ij} = \bar{r} + b_i^{(U)} + b_j^{(I)} +
        \boldsymbol{q}_j^T \cdot \left(\boldsymbol{p}_i + |R^{(U)}(i)|^{-1/2} \sum_{g \in R^{(U)}(i)}\boldsymbol{y}_g\right),
\end{dmath}
where $R^{(U)}(i)$ returns the implicit feedback (i.e., the set of rated items)
from user $i$, and $\boldsymbol{y}_g \in R^{k \times 1}$ denotes the latent
factors of item $g$.  The idea of incorporating the implicit feedback is that
users' implicit feedback on the items (e.g., browsed an item or rated an item)
should also reflect her preferences on the target item $j$.  The interaction
between the implicit feedback and the target item $j$ is again modeled by the
inner product of their corresponding latent factors.  Since the number of a
user's implicit feedback is often much larger, SVD++ introduces a normalization
term $|R^{(U)}(i)|^{-1/2}$ so that the latent factors of the implicit feedback
will not dominate the result.

\subsection{The Weighted-SVD model} \label{sec:wsvd-model}

The SVD model essentially assumes that each latent factor has equal weight,
which may not always be a reasonable assumption.  For example, to predict
users' ratings on the movies, the latent factor that (implicitly) represents
the genre of a movie is probably more important than the latent factor that
represents the number of actors/actresses in the movie.  Unfortunately, the
simple inner product operation in Equation~\ref{eq:latent-model} and
Equation~\ref{eq:svd++-model} cannot assign different weights to different
latent factors.  This motivates the Weighted-SVD (WSVD) model, which introduces
the weights of each latent factor into the model, as shown in
Equation~\ref{eq:wsvd-model}.

\begin{dmath} \label{eq:wsvd-model}
\hat{r}_{ij} = \bar{r} + b_i^{(U)} + b_j^{(I)} + (\boldsymbol{w} \odot \boldsymbol{p}_i)^T \cdot \boldsymbol{q}_j,
\end{dmath}
where $\boldsymbol{w} \in R^{k \times 1}$ is the vector of weights of the latent
factors, and the $\odot$ operator denotes the Hadamard product (i.e.,
element-wise multiplication) on the two column vectors $\boldsymbol{w}$ and
$\boldsymbol{p}_i$.

Compared with the SVD model, in the Weighted-SVD model each latent
factor is multiplied by a weight.  Thus, different latent factors have distinct
weights to influence user $i$'s rating on item $j$.

\begin{algorithm}[tb] \label{alg:sgd}
    \SetAlgoLined
    \KwData{The rated (user, item) pairs $\mathcal{K}$,
            the regularization coefficients $\boldsymbol\lambda = (\lambda_w, \lambda_p, \lambda_q, \lambda_U, \lambda_I)$,
            the learning rates $\boldsymbol{\eta} = (\eta_w, \eta_p, \eta_q, \eta_U, \eta_I)$,
            the learning decay rate $\gamma$}
    \KwResult{Model parameters $\boldsymbol{\Theta} = (\boldsymbol{w}, \boldsymbol{P}, \boldsymbol{Q}, \boldsymbol{b}^{(U)}, \boldsymbol{b}^{(I)})$}
    $\boldsymbol{P} \leftarrow \mathcal{N}(\boldsymbol{0},\boldsymbol{1})$; $\boldsymbol{Q} \leftarrow \mathcal{N}(\boldsymbol{0},\boldsymbol{1})$; $\boldsymbol{w} \leftarrow \boldsymbol{1}$; $\boldsymbol{b}^{(U)} \leftarrow \boldsymbol{0}$; $\boldsymbol{b}^{(I)} \leftarrow \boldsymbol{0}$; $epoch \leftarrow 0$\; 
    \Repeat{termination condition is met}{
        \For{$(i,j) \in \mathcal{K}$} {
            $b_i^{(U)} \leftarrow b_i^{(U)} - \gamma^{epoch} \eta_U \frac{\partial \mathcal{L}(\boldsymbol{\Theta})}{\partial b_i^{(U)}}$ (based on Equation~\ref{eq:wsvd-deri-bu})\;
            $b_j^{(I)} \leftarrow b_j^{(I)} - \gamma^{epoch} \eta_I \frac{\partial \mathcal{L}(\boldsymbol{\Theta})}{\partial b_j^{(I)}}$ (based on Equation~\ref{eq:wsvd-deri-bi})\;
            $\boldsymbol{w} \leftarrow \boldsymbol{w} - \gamma^{epoch} \eta_w \frac{\partial \mathcal{L}(\boldsymbol{\Theta})}{\partial \boldsymbol{w}}$ (based on Equation~\ref{eq:wsvd-deri-w})\;
            $\boldsymbol{p}_i \leftarrow \boldsymbol{p}_i - \gamma^{epoch} \eta_p \frac{\partial \mathcal{L}(\boldsymbol{\Theta})}{\partial \boldsymbol{p}_i}$ (based on Equation~\ref{eq:wsvd-deri-p})\;
            $\boldsymbol{q}_j \leftarrow \boldsymbol{q}_j - \gamma^{epoch} \eta_q \frac{\partial \mathcal{L}(\boldsymbol{\Theta})}{\partial \boldsymbol{q}_j}$ (based on Equation~\ref{eq:wsvd-deri-q})\;
        }
        $epoch \leftarrow epoch + 1$\;
    }
    \caption {The Weighted-SVD learning model based on SGD}
\end{algorithm}

\subsubsection{Learning the Weighted-SVD model}

The learning process can be modeled as an optimization problem: obtaining the
parameters $\boldsymbol{\Theta}$ that minimize the loss function $\mathcal{L}$,
which can be defined by Equation~\ref{eq:wsvd-loss-func}.

\begin{dmath} \label{eq:wsvd-loss-func}
\mathcal{L}(\boldsymbol{\Theta}) \equiv \frac{1}{2} \sum_{(i,j) \in \mathcal{K}} (r_{ij} - \hat{r}_{ij})^2 + \frac{\boldsymbol{\lambda}}{2} \norm{\boldsymbol{\Theta}}_2 \\
= \frac{1}{2} \sum_{(i,j) \in \mathcal{K}} \left(r_{ij} - \left(\bar{r}+b_i^{(U)} + b_j^{(I)} + (\boldsymbol{w} \odot \boldsymbol{p}_i)^T \cdot \boldsymbol{q}_j \right)\right)^2 + \left(\frac{\lambda_w}{2}\norm{\boldsymbol{w}}_2 + \frac{\lambda_p}{2}\norm{\boldsymbol{P}}_2 + \frac{\lambda_q}{2}\norm{\boldsymbol{Q}}_2 + \frac{\lambda_U}{2}\norm{\boldsymbol{b}^{(U)}}_2 + \frac{\lambda_I}{2}\norm{\boldsymbol{b}^{(I)}}_2\right),
\end{dmath}
where $r_{ij}$ is the real rating of user $i$ on item $j$,
$\boldsymbol{P} = \left[\boldsymbol{p}_1, \boldsymbol{p}_2, \ldots, \boldsymbol{p}_m\right]^T \in R^{m \times k}$,
$\boldsymbol{Q} = \left[\boldsymbol{q}_1, \boldsymbol{q}_2, \ldots, \boldsymbol{q}_n\right] \in R^{k \times n}$,
$\boldsymbol{b}^{(U)} = \left[b_1^{(U)}, b_2^{(U)}, \ldots, b_m^{(U)}\right]^T \in R^{m \times 1}$,
$\boldsymbol{b}^{(I)} = \left[b_1^{(I)}, b_2^{(I)}, \ldots, b_n^{(I)}\right]^T \in R^{n \times 1}$,
$\boldsymbol{\Theta} = (\boldsymbol{w}, \boldsymbol{P}, \boldsymbol{Q}, \boldsymbol{b}^{(U)}, \boldsymbol{b}^{(I)})$
is the set of unknown parameters to learn,
$\boldsymbol{\lambda} = (\lambda_w, \lambda_p, \lambda_q, \lambda_U, \lambda_I)$
is the set of the regularization coefficients,
$m$ and $n$ are the number of users and items respectively, and
$k$ is the pre-defined number of latent factors. 
The Frobenius norm ($\norm{\cdot}_2$) is used on the parameters for regularization.

We apply the stochastic gradient descent (SGD) approach to obtain the
parameters.  Let $\boldsymbol{w} = [w_1, w_2, \ldots, w_k]^T$, $\boldsymbol{p}_i =
[p_{i1}, p_{i2}, \ldots, p_{ik}]^T$, and $\boldsymbol{q}_j = [q_{1j}, q_{2j},
\ldots, q_{kj}]^T$, the partial derivatives of the loss function with respect
to the parameters based on user $i$'s rating on item $j$ are given in
Equation~\ref{eq:wsvd-deri-bu} to Equation~\ref{eq:wsvd-deri-q}.

\begin{dmath} \label{eq:wsvd-deri-bu}
\frac{\partial \mathcal{L}(\boldsymbol{\Theta})}{\partial b_i^{(U)}} = -\left(r_{ij} - \bar{r} - b_i^{(U)} - b_j^{(I)} - (\boldsymbol{w} \odot \boldsymbol{p}_i)^T \cdot \boldsymbol{q}_j \right) + \lambda_U b_i^{(U)}
\end{dmath}

\begin{dmath} \label{eq:wsvd-deri-bi}
\frac{\partial \mathcal{L}(\boldsymbol{\Theta})}{\partial b_j^{(I)}} = -\left(r_{ij} - \bar{r} - b_i^{(U)} - b_j^{(I)} - (\boldsymbol{w} \odot \boldsymbol{p}_i)^T \cdot \boldsymbol{q}_j \right) + \lambda_I b_j^{(I)}
\end{dmath}

\begin{dmath} \label{eq:wsvd-deri-w}
\frac{\partial \mathcal{L}(\boldsymbol{\Theta})}{\partial \boldsymbol{w}} = -\left(r_{ij} - \bar{r} - b_i^{(U)} - b_j^{(I)} - (\boldsymbol{w} \odot \boldsymbol{p}_i)^T \cdot \boldsymbol{q}_j \right) (\boldsymbol{p}_i \odot \boldsymbol{q}_j) + \lambda_w \boldsymbol{w}
\end{dmath}

\begin{dmath} \label{eq:wsvd-deri-p}
\frac{\partial \mathcal{L}(\boldsymbol{\Theta})}{\partial \boldsymbol{p}_i} = -\left(r_{ij} - \bar{r} - b_i^{(U)} - b_j^{(I)} - (\boldsymbol{w} \odot \boldsymbol{p}_i)^T \cdot \boldsymbol{q}_j \right) (\boldsymbol{w} \odot \boldsymbol{q}_j) + \lambda_p \boldsymbol{p}_i
\end{dmath}

\begin{dmath} \label{eq:wsvd-deri-q}
\frac{\partial \mathcal{L}(\boldsymbol{\Theta})}{\partial \boldsymbol{q}_j} = -\left(r_{ij} - \bar{r} - b_i^{(U)} - b_j^{(I)} - (\boldsymbol{w} \odot \boldsymbol{p}_i)^T \cdot \boldsymbol{q}_j \right) (\boldsymbol{w} \odot \boldsymbol{p}_i) + \lambda_q \boldsymbol{q}_j
\end{dmath}

Algorithm~\ref{alg:sgd} shows the training algorithm based on SGD.  The
algorithm iterates over each (user, item) pair in $\mathcal{K}$ and updates the
parameters for multiple epochs until the termination condition (e.g., the
parameters are converged, or the number of epochs reaches a specified
threshold) is met.

\subsubsection{Analysis of the Weighted-SVD model and the related methods} \label{sec:model-analysis}

This section analyzes the number of learnable parameters of the WSVD, SVD,
SVD++, and the PMF model along with their training costs.

For the PMF model, each user $i$ is associated with a vector of latent factors
$\boldsymbol{p}_i$ of size $k$, and each item $j$ is associated with a vector
of latent factors $\boldsymbol{q}_j$ of size $k$.  Thus, the total number of
learnable parameters is $mk + nk$.

In the SVD model, in addition to the vectors of latent factors for all the
users and all the items, each user $i$ also has a user bias $b_i^{(U)}$, and
each item $j$ has an item bias $b_j^{(I)}$.  Therefore, the total number of
parameters needs to learn is $m(k+1) + n(k+1)$.

In the Weighted-SVD model, in addition to the above parameters, each latent
factor is also associated with a weight parameter.  Therefore, the number of
parameters need to learn becomes $m(k+1) + n(k+1) + k$.  Since $k \ll m$ and $k
\ll n$ in most settings, the parameters to learn in the Weighted-SVD model is
only slightly larger than the original SVD model. 

Another related model, SVD++, requires to learn not only $b_i^{(U)}$,
$b_j^{(I)}$, $\boldsymbol{p}_i$, and $\boldsymbol{q}_j$ for all user $i$ and
item $j$, but also $\boldsymbol{y}_g$ the latent factors of the item $g$ that
the user $i$ had rated (i.e., the implicit feedback).  Even if each item only
associates with one type of implicit feedback, the total number of parameters
needs to learn is $m(k+1) + n(k+1) + nk$, which is much larger than the number
of parameters in the SVD and the Weighted-SVD model.  As a result, SVD++ may
require a much larger dataset or stronger regularization terms to prevent
overfitting.

We use Algorithm~\ref{alg:sgd} to explain the required training time of the
WSVD model.  The parameter updating procedure, based on one observed rating
$r_{ij}$, is shown from line 4 to line 8.  Since the number of latent factors
is usually not large, typically dozens to hundreds, the processing time of each
parameter update is short.  In addition, the algorithm typically takes only
dozens of epochs for the parameters to converge.  As a result, the required
training time of the WSVD model grows linearly with the number of known
ratings.  For a similar reason, the training time of the SVD and the PMF models
also grows linearly with the training data size.  However, the running time of
the SVD++ model grows super-linearly with the size of the dataset.  This is
because for each rating $r_{ij}$ the SVD++ model needs to update not only the
biases and the latent factors of user $i$ and item $j$, but also the latent
factors of user $i$'s rated items.  As more ratings are available, the number
of the items rated by user $i$ may increase super-linearly.  As a result, the
computation time would dramatically rise.

\section{Experiments} \label{sec:exp}

\subsection{Experimental data and settings}

\begin{table}[tb]
\centering
\caption{Statistics of the experimental datasets}
\label{tab:exp-data-stats}
\begin{tabular}{c||r|r|r|r|c}
\hline
Dataset        & \# users & \# items & \# ratings & Density  & Rating Scale                         \\ \hline\hline
MovieLens-100K & 943      & 1,682    & 100,000    & 6.3047\% & {[}1, 2, 3, 4, 5{]}                  \\ \hline
MovieLens-1M   & 6,040    & 3,704    & 1,000,209  & 4.4684\% & {[}1, 2, 3, 4, 5{]}                  \\ \hline
MovieLens-10M  & 69,878   & 10,677   & 10,000,054 & 1.3403\% & {[}1, 2, 3, 4, 5{]}                  \\ \hline
FilmTrust      & 1,508    & 2,071    & 35,497     & 1.1366\% & {[}0.5, 1, 1.5, 2, 2.5, 3, 3.5, 4{]} \\ \hline
Epinions       & 40,163   & 139,738  & 664,824    & 0.0118\% & {[}1, 2, 3, 4, 5{]}                  \\ \hline
\end{tabular}
\end{table}

We compared the Weighted-SVD model with several Matrix Factorization methods,
including the PMF, SVD, and the SVD++ models~\cite{mnih2008probabilistic,
koren2008factorization, koren2009matrix}.
We used the hyper-parameters suggested in~\cite{ricci2015recommender}: for SVD,
the initial learning rates are set to 0.005 and the regularization terms are
set to 0.02; for SVD++, the initial learning rates are set to 0.007 and the
regularization terms for the biases, latent factors, and implicit latent
factors are set to 0.005, 0.015, and 0.015 respectively.  The hyper-parameters
for WSVD and PMF are the same as SVD.  The learning decay rates, if required,
are set to 0.9 in all models.  We set the number of latent factors to 15.

We used the following datasets as experimental data: (1)
MovieLens-100K~\cite{harper2016movielens}, (2)
MovieLens-1M~\cite{harper2016movielens}, (3)
MovieLens-10M~\cite{harper2016movielens}, (4)
FilmTrust~\cite{guo2013novel}, (5) Epinions~\cite{massa2007trust}.
Statistics of these datasets are shown in Table~\ref{tab:exp-data-stats}.  For
each dataset, we took $80\%$ of the ratings as the training data and the rest
$20\%$ of the ratings as the test data.

\subsection{Comparing the prediction errors}

\begin{table}[tb]
\centering
\caption{A comparison of the RMSEs of the latent factor models on several datasets}
\label{tab:model-rmses}
\begin{tabular}{l|l||l|l|l|l}
\hline
                                &          & WSVD       & SVD    & SVD++      & PMF    \\ \hline\hline
\multirow{2}{*}{MovieLens-100K} & Training & 0.9246     & 0.8912 & \bf{0.8579}& 0.9326 \\ \cline{2-6} 
                                & Test     & \bf{0.9430}& 1.3076 & 1.2622     & 1.4109 \\ \hline
\multirow{2}{*}{MovieLens-1M}   & Training & 0.8964     & 0.8739 & \bf{0.8600}& 0.8761 \\ \cline{2-6} 
                                & Test     & \bf{0.9921}& 0.9927 & 0.9947     & 3.7888 \\ \hline
\multirow{2}{*}{MovieLens-10M}  & Training & 0.8569     & 0.8532 & \bf{0.8498}& 0.8530 \\ \cline{2-6} 
                                & Test     & \bf{0.9470}& 0.9472 & 0.9486     & 3.6673 \\ \hline
\multirow{2}{*}{FilmTrust}      & Training & 0.7704     & 0.7635 & \bf{0.7194}& 1.1306 \\ \cline{2-6} 
                                & Test     & \bf{0.8899}& 0.8935 & 0.8947     & 3.1617 \\ \hline
\multirow{2}{*}{Epinions}       & Training & 0.9970     & 0.9654 & \bf{0.8468}& 2.2092 \\ \cline{2-6} 
                                & Test     & 1.0928 & 1.1072 & 1.1037     & 4.2170 \\ \hline
\end{tabular}
\end{table}

\figtwo{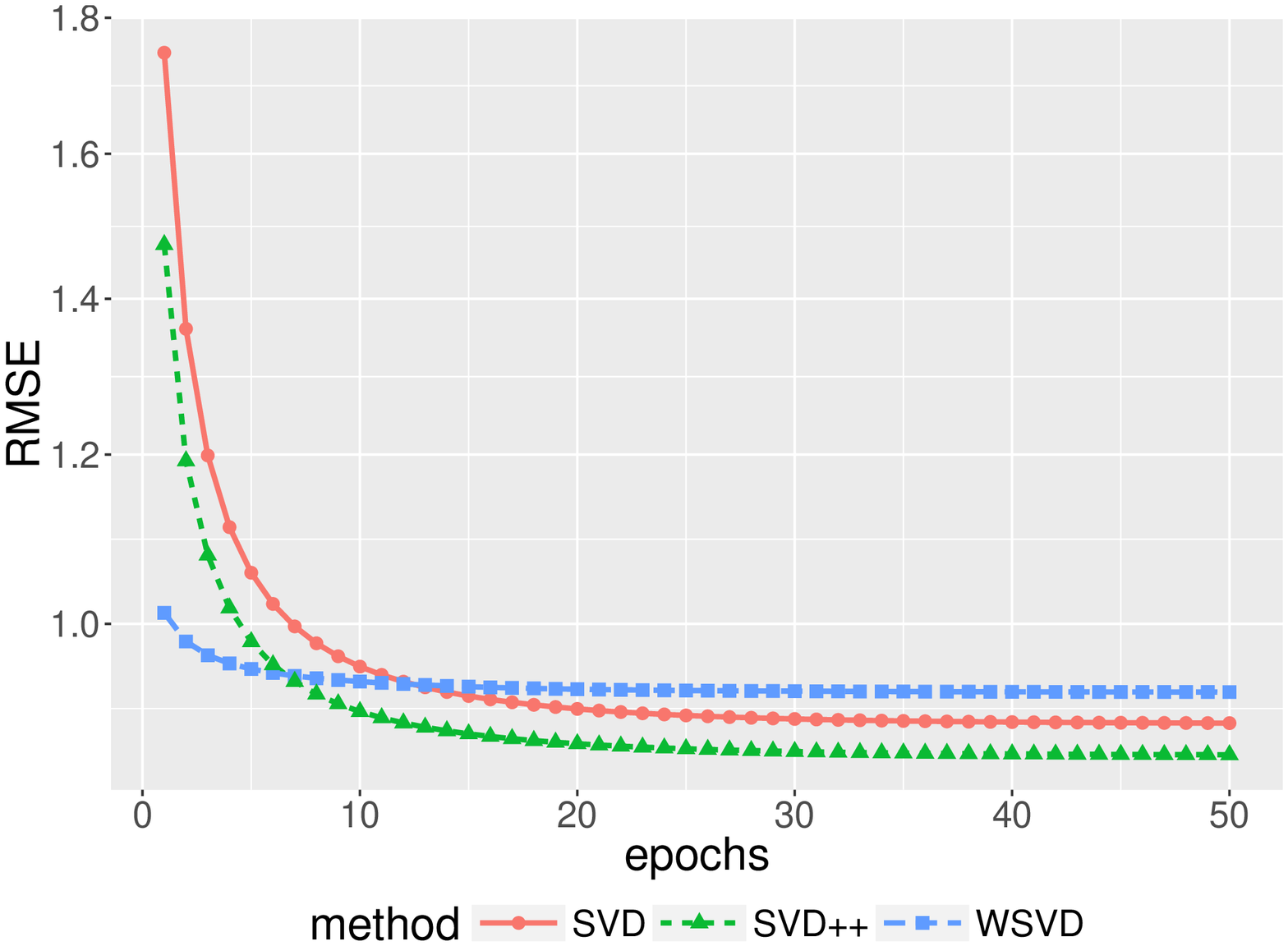}{The RMSE scores of the training data.}{fig:ml-100k-train-rmse-to-epoch}
       {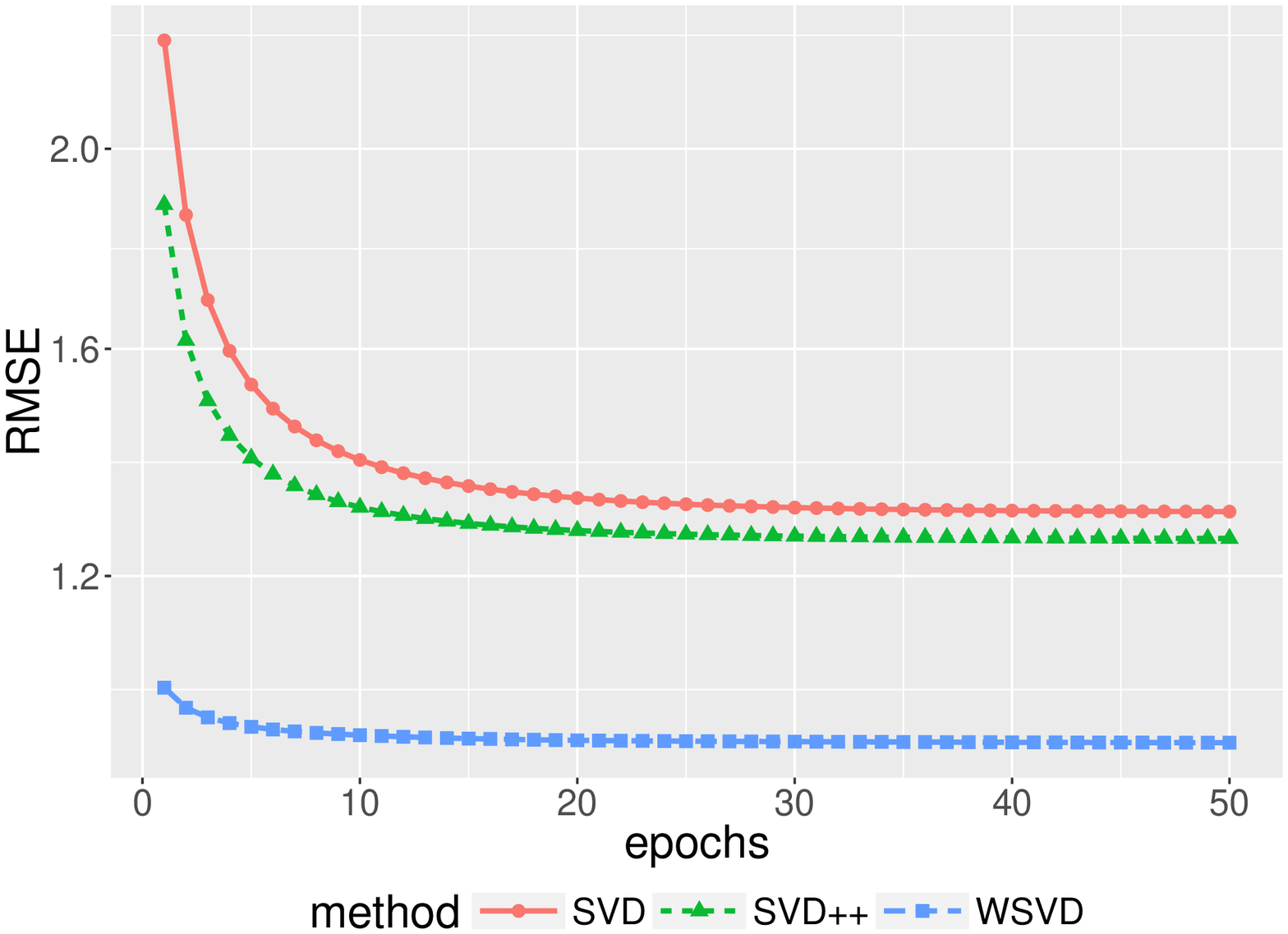}{The RMSE scores of the test data.}{fig:ml-100k-test-rmse-to-epoch}
       {The relationship between the RMSE scores and the epochs using the MovieLens-100K data.}{fig:ml-100k-rmse-to-epoch}

\figtwo{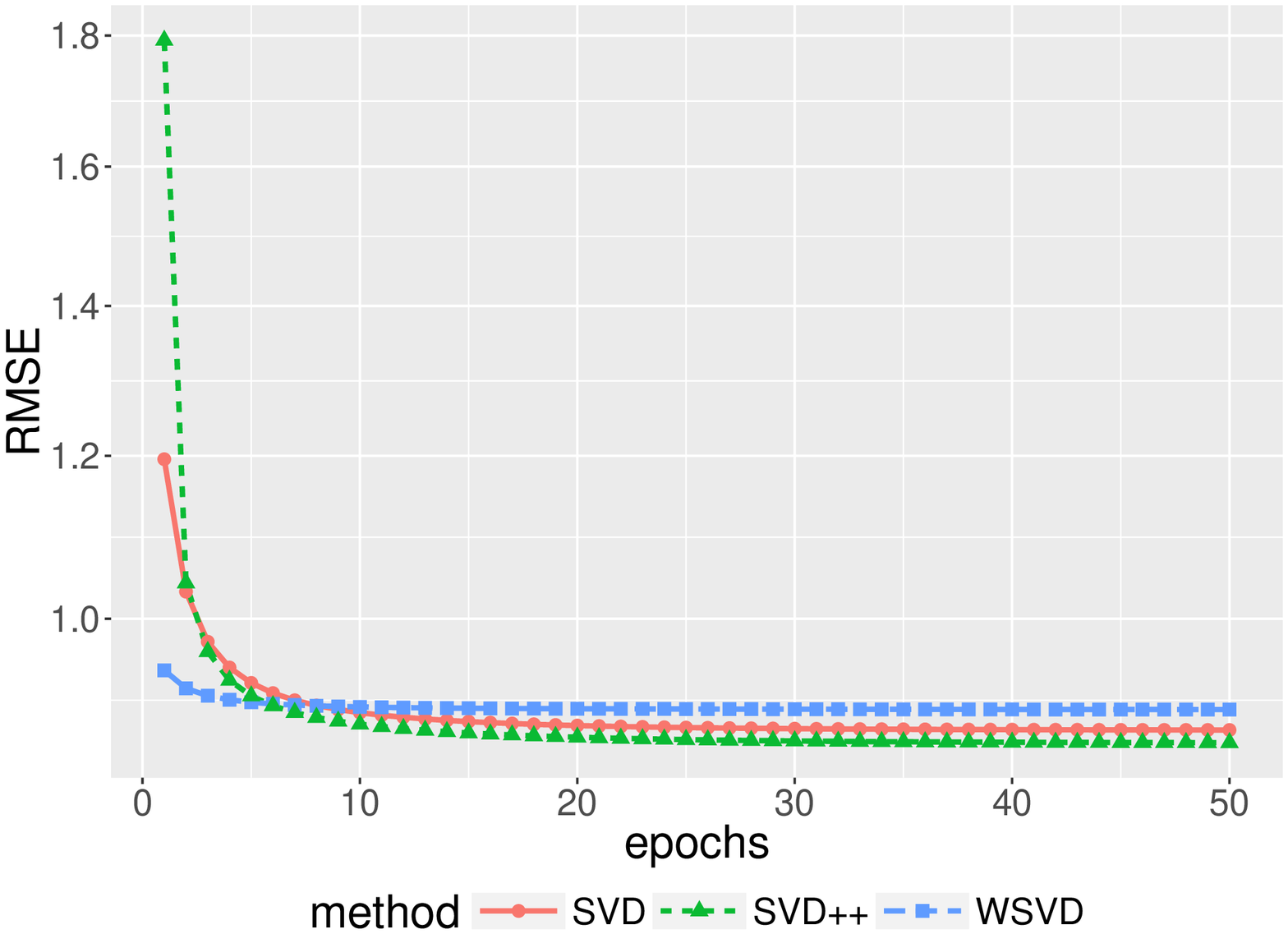}{The RMSE scores of the training data.}{fig:ml-1m-train-rmse-to-epoch}
       {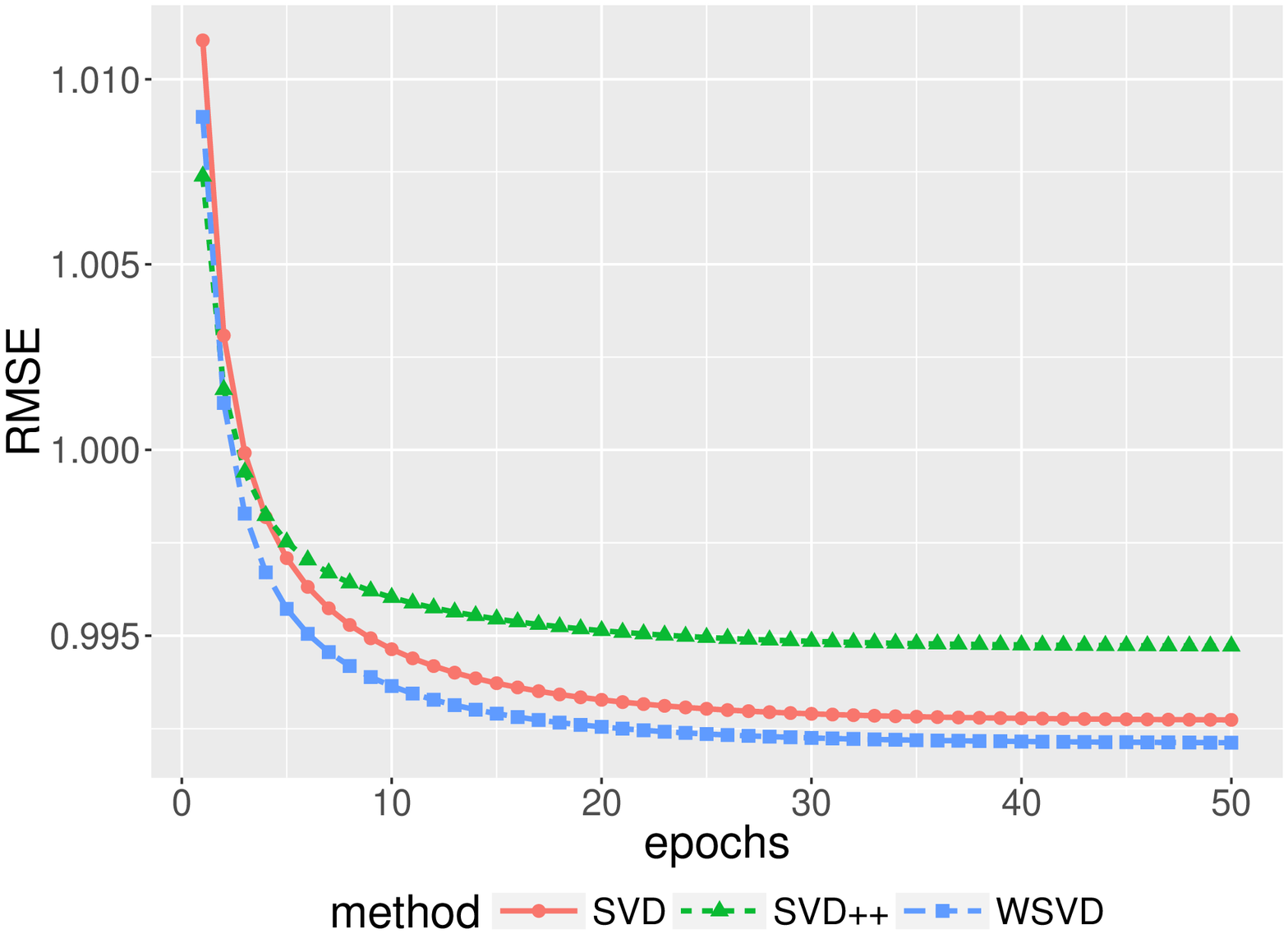}{The RMSE scores of the test data.}{fig:ml-1m-test-rmse-to-epoch}
       {The relationship between the RMSE scores and the epochs using the MovieLens-1M data.}{fig:ml-1m-rmse-to-epoch}

\figtwo{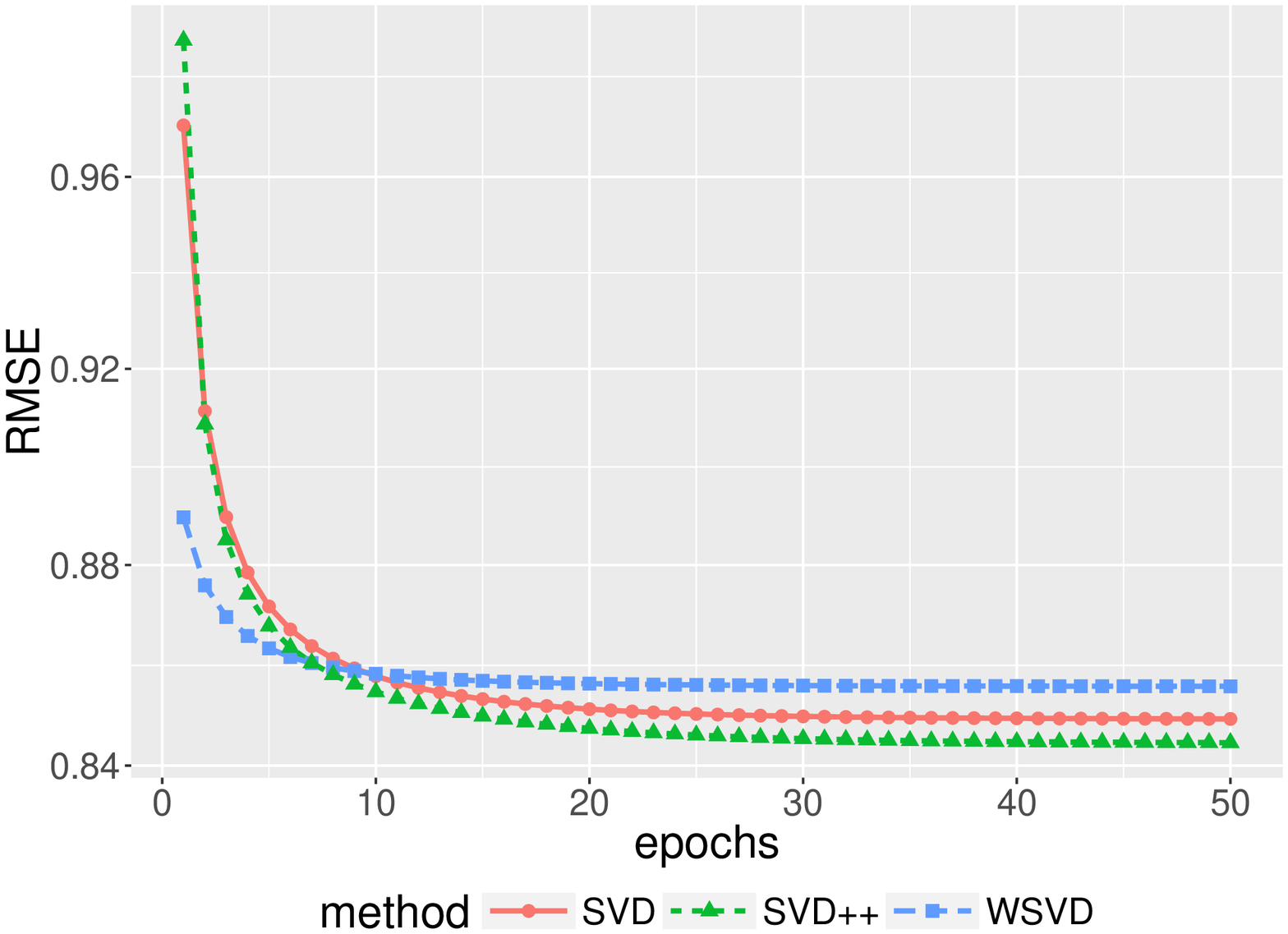}{The RMSE scores of the training data.}{fig:ml-10m-train-rmse-to-epoch}
       {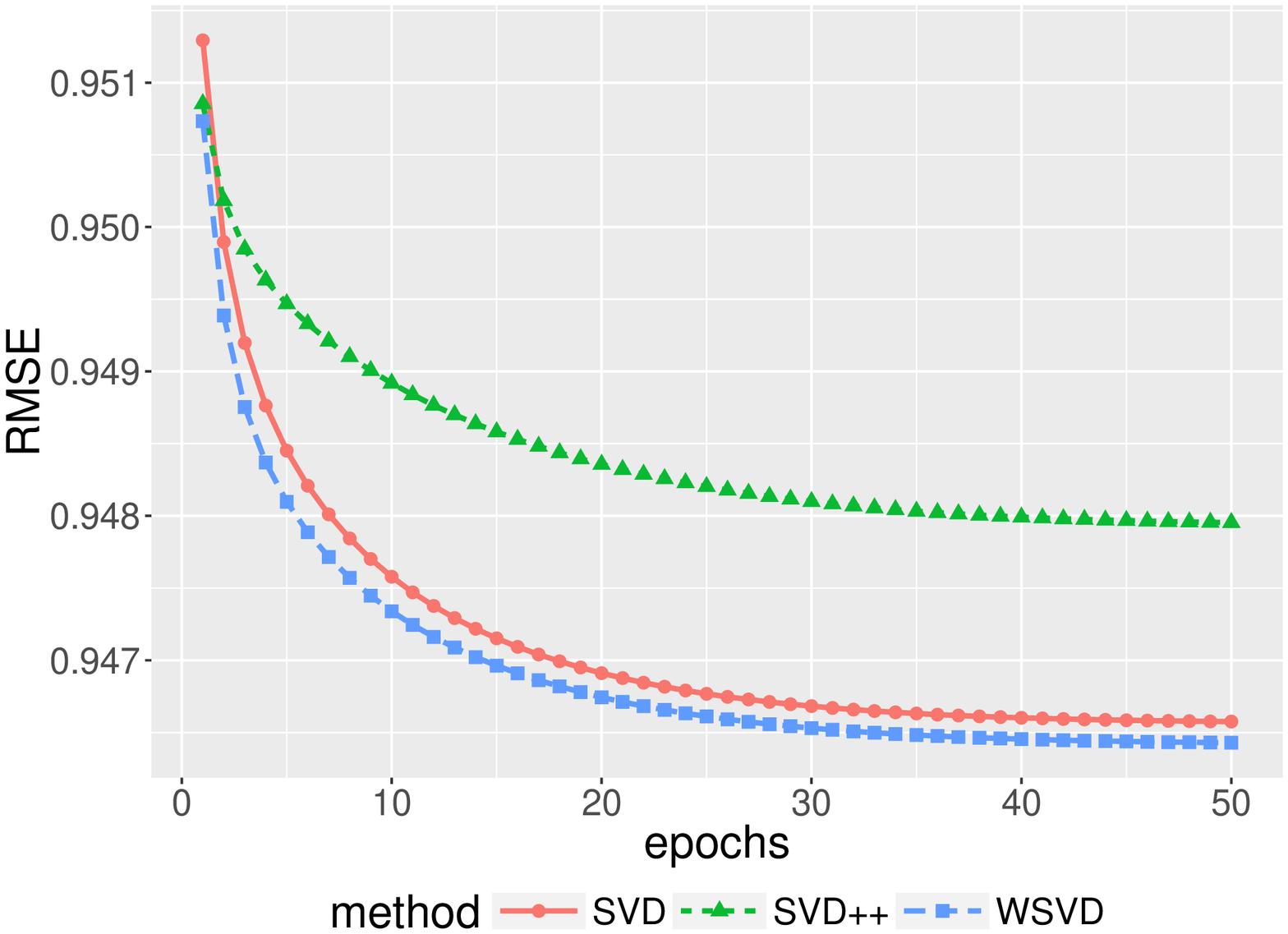}{The RMSE scores of the test data.}{fig:ml-10m-test-rmse-to-epoch}
       {The relationship between the RMSE scores and the epochs using the MovieLens-10M data.}{fig:ml-10m-rmse-to-epoch}

\figtwo{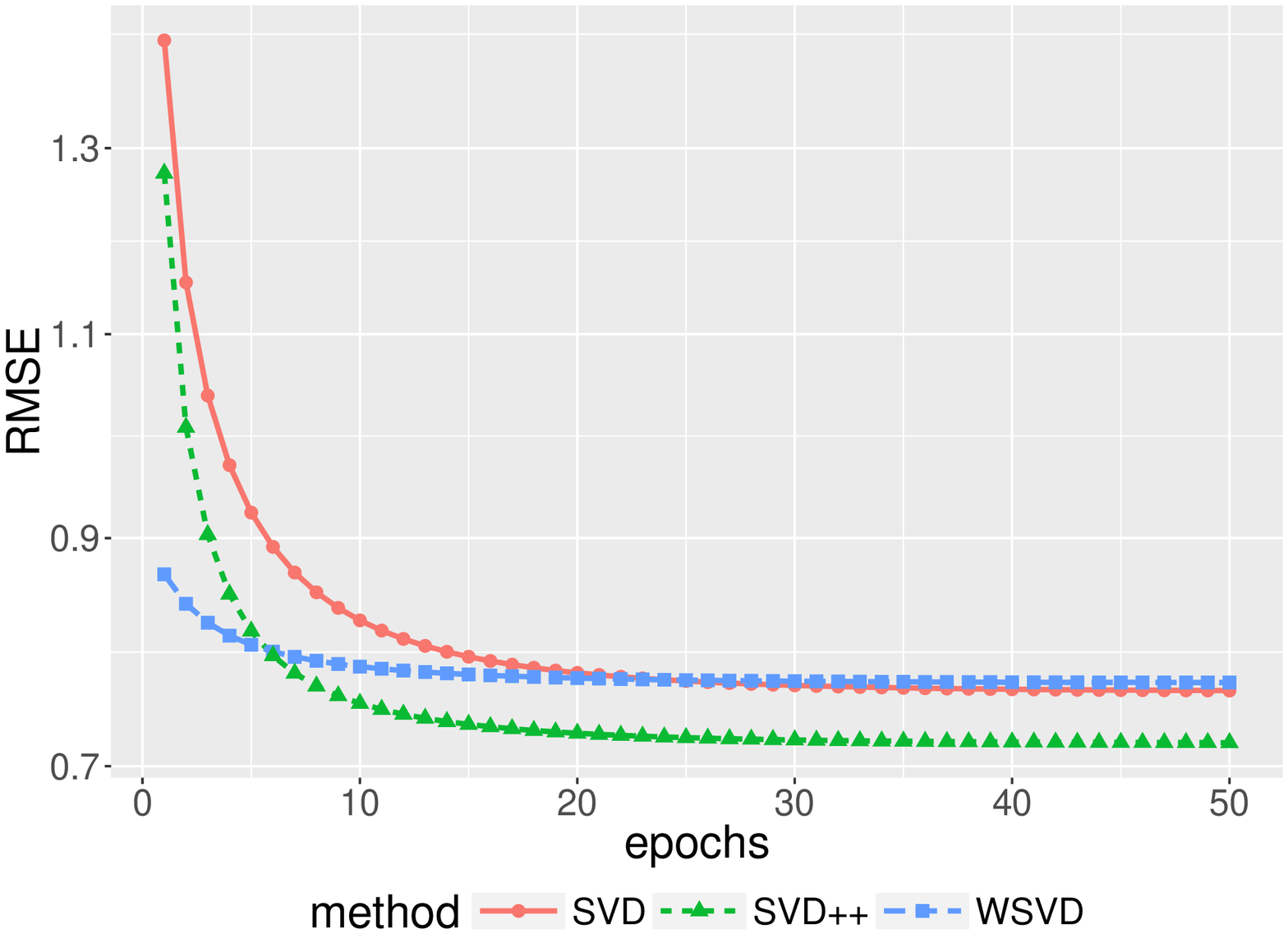}{The RMSE scores of the training data.}{fig:filmtrust-train-rmse-to-epoch}
       {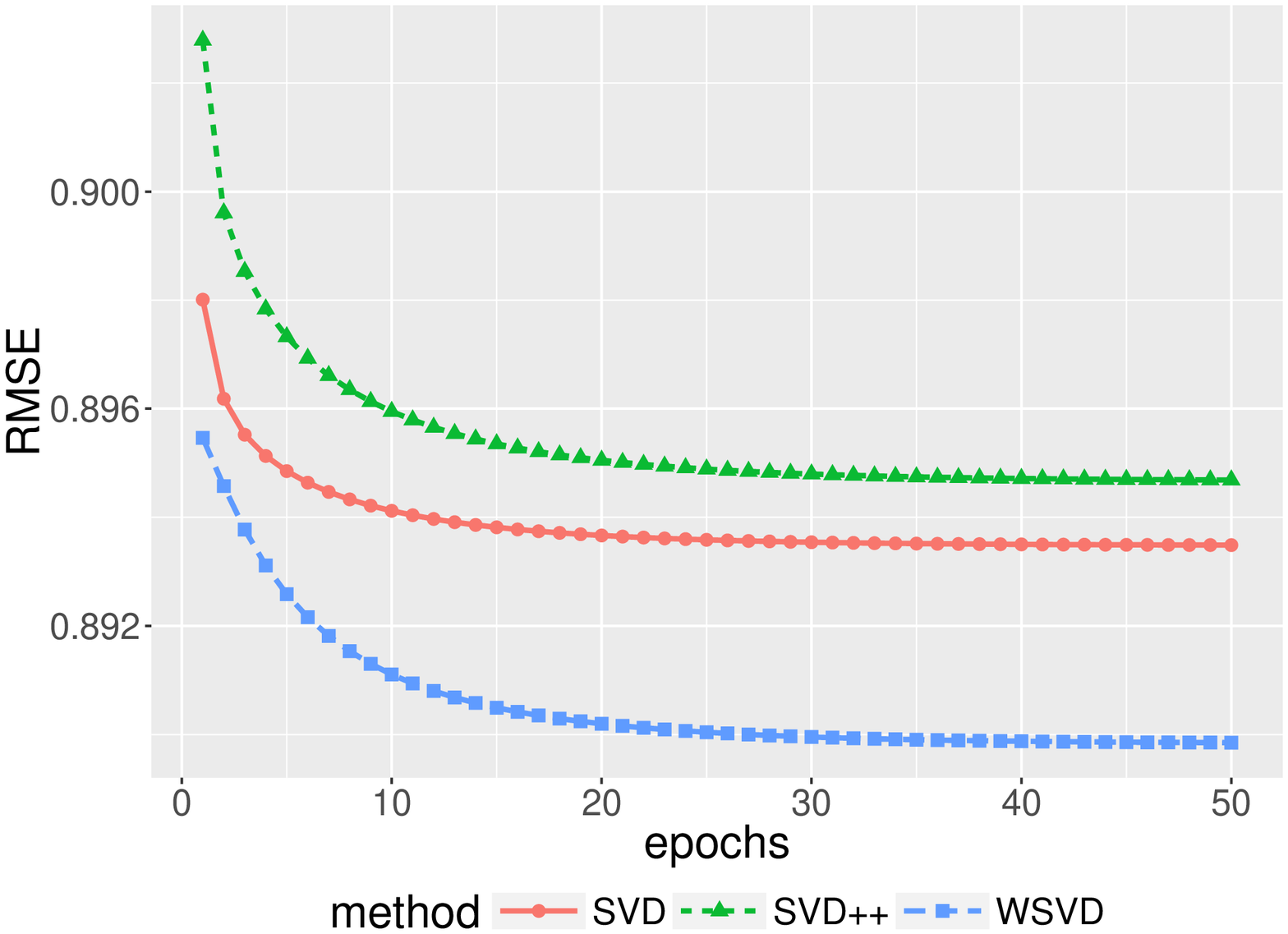}{The RMSE scores of the test data.}{fig:filmtrust-test-rmse-to-epoch}
       {The relationship between the RMSE scores and the epochs using the FilmTrust data.}{fig:filmtrust-rmse-to-epoch}

\figtwo{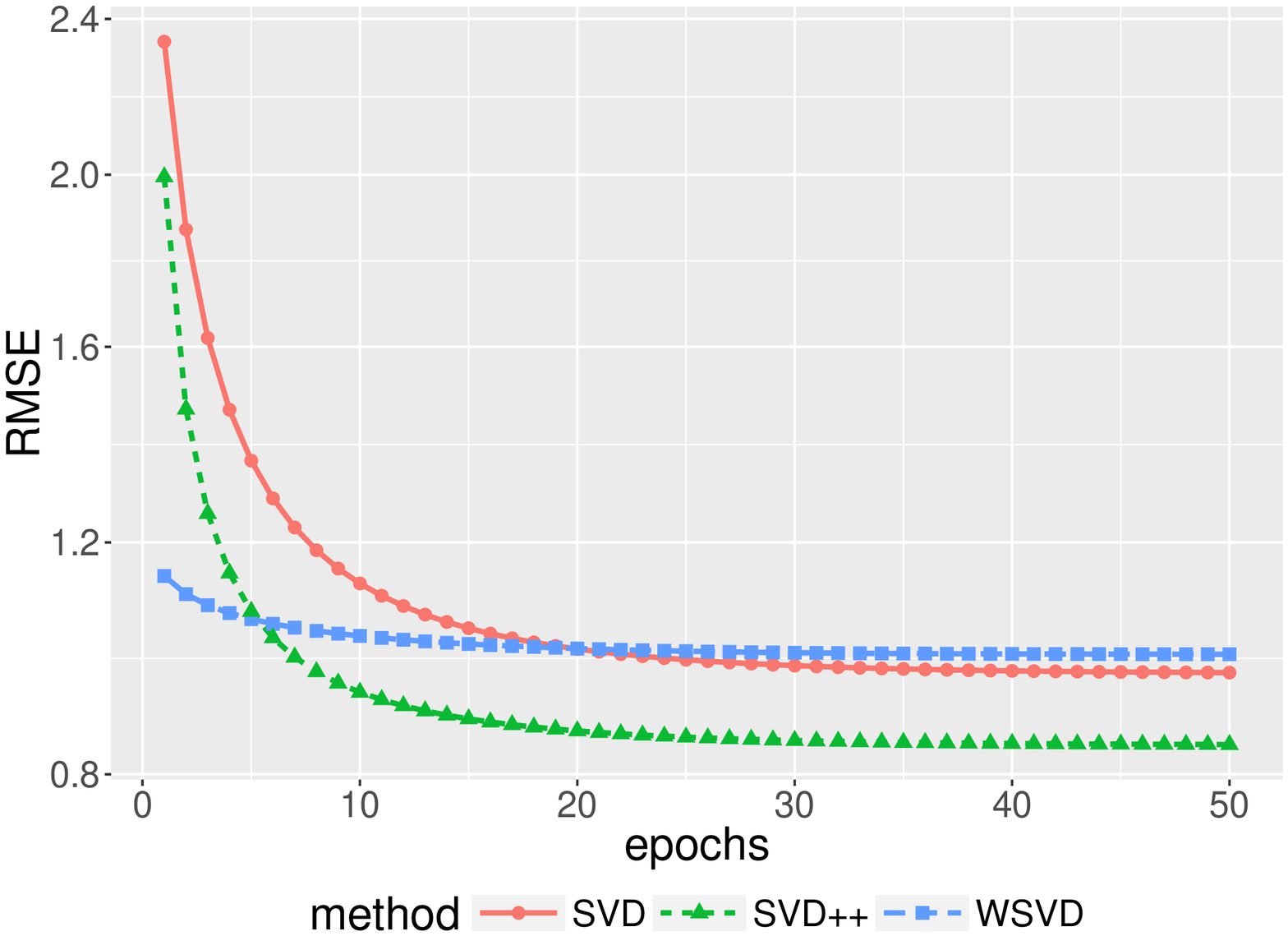}{The RMSE scores of the training data.}{fig:epinions-train-rmse-to-epoch}
       {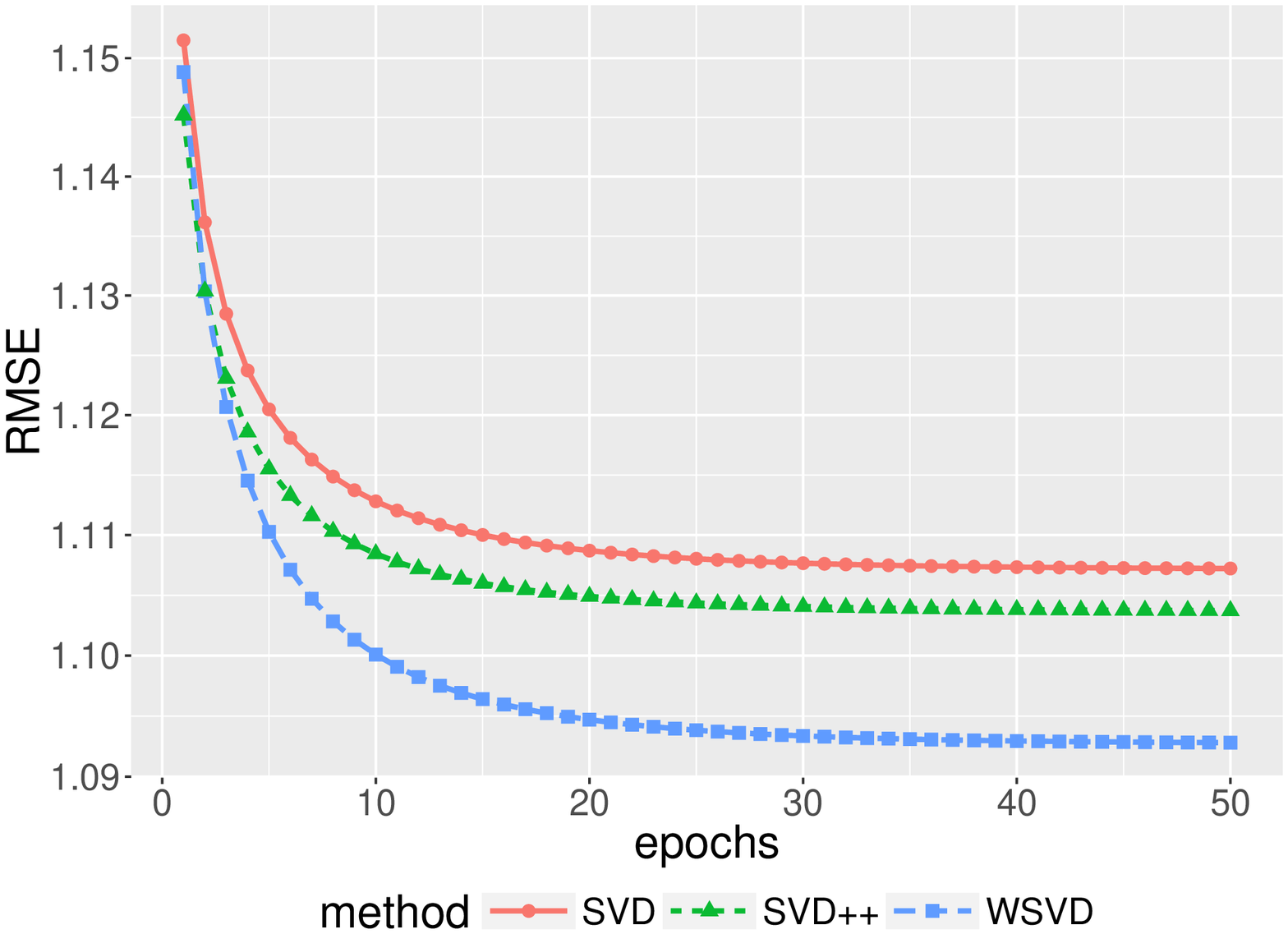}{The RMSE scores of the test data.}{fig:epinions-test-rmse-to-epoch}
       {The relationship between the RMSE scores and the epochs using the Epinions data.}{fig:epinions-rmse-to-epoch}

We used the RMSE scores to compare different methods.  The results are shown in
Table~\ref{tab:model-rmses}.  As highlighted, the SVD++ model performs best
(i.e., the lowest RMSE score) for each of the training dataset.  This is not
surprising since the SVD++ model contains much more learnable parameters.
%However, when we compare the RMSE based on the test data, the WSVD model
%outperforms all the baselines.  
Compared to the SVD model, the WSVD model slightly increases the number of
learnable parameters.  However, the test RMSE of the WSVD model is lower than
the other models in all the experimented datasets.  This suggests that
assigning different weight values to different latent factors probably make the
model closer to the real scenario.  The performance of the PMF model is
acceptable in the training data but much worse on the test data.  This is
probably because the PMF model does not include the average ratings, the user
bias, and the test bias, which could be critical clues in the preference
prediction task.

Figure~\ref{fig:ml-100k-rmse-to-epoch}, Figure~\ref{fig:ml-1m-rmse-to-epoch},
Figure~\ref{fig:ml-10m-rmse-to-epoch}, Figure~\ref{fig:filmtrust-rmse-to-epoch}
and Figure~\ref{fig:epinions-rmse-to-epoch} show the training and the test RMSE
scores of the compared methods on the five experimental datasets as the number
of epoch increases.  As shown, the training and the test RMSE scores of all the
methods on all the datasets gradually decrease and converge.  In most cases,
the RMSE scores converges in 10 to 20 epochs.  Note that we do not include the
performance of the PMF model in the figures, since it performs much worse than
the other methods, as shown in Table~\ref{tab:model-rmses}.  The WSVD model
performs best in terms of the test RMSE score almost through the entire process
on all the experimental datasets.

\subsection{Comparing different hyper-parameters} \label{sec:diff-hyper-para}

\begin{table}[tb]
\centering
\caption{Performance comparison with different hyper-parameters (using the Movielens-100K dataset)}
\label{tab:hyper-para-cmp}
\begin{tabular}{l|l||l|l|l|l}
\hline
\multicolumn{2}{l||}{Hyper-parameters} & \multicolumn{4}{l}{Test RMSE scores} \\ \hline\hline
$k$ & $\lambda$ & WSVD  & SVD   & SVD++ & PMF   \\ \hline\hline
10 & 0.001 & \bf{0.9903} & 1.6765 & 1.6898 & 3.0223\\ \hline
10 & 0.005 & \bf{0.9909} & 1.6852 & 1.6404 & 3.0990\\ \hline
10 & 0.01  & \bf{0.9904} & 1.6672 & 1.6264 & 3.0204\\ \hline
10 & 0.05  & \bf{0.9918} & 1.6400 & 1.6225 & 3.0919\\ \hline
10 & 0.1   & \bf{0.9924} & 1.5816 & 1.5724 & 3.0098\\ \hline
10 & 0.5   & \bf{1.0038} & 1.3315 & 1.3187 & 3.1695\\ \hline
10 & 1     & \bf{1.0171} & 1.1735 & 1.1495 & 3.3384\\ \hline
20 & 0.001 & \bf{0.9919} & 2.2516 & 2.3030 & 3.4681\\ \hline
20 & 0.005 & \bf{0.9912} & 2.2248 & 2.1572 & 3.2964\\ \hline
20 & 0.01  & \bf{0.9920} & 2.2352 & 2.1709 & 3.4595\\ \hline
20 & 0.05  & \bf{0.9920} & 2.1471 & 2.0970 & 3.2526\\ \hline
20 & 0.1   & \bf{0.9936} & 2.0836 & 2.0656 & 3.3871\\ \hline
20 & 0.5   & \bf{1.0038} & 1.6077 & 1.5892 & 3.0832\\ \hline
20 & 1     & \bf{1.0172} & 1.3249 & 1.3073 & 3.4214\\ \hline
40 & 0.001 & \bf{0.9915} & 3.1927 & 3.3221 & 4.0514\\ \hline
40 & 0.005 & \bf{0.9938} & 3.1454 & 3.1644 & 4.0885\\ \hline
40 & 0.01  & \bf{0.9917} & 3.1646 & 3.0714 & 4.0278\\ \hline
40 & 0.05  & \bf{0.9941} & 3.0098 & 2.9805 & 3.9757\\ \hline
40 & 0.1   & \bf{0.9933} & 2.9046 & 2.8642 & 3.8154\\ \hline
40 & 0.5   & \bf{1.0041} & 2.0719 & 2.0820 & 3.3267\\ \hline
40 & 1     & \bf{1.0172} & 1.6053 & 1.5535 & 3.2051\\ \hline
80 & 0.001 & \bf{0.9946} & 5.0289 & 5.5995 & 5.6352\\ \hline
80 & 0.005 & \bf{0.9957} & 4.9873 & 5.1369 & 5.7221\\ \hline
80 & 0.01  & \bf{0.9945} & 4.9739 & 4.9098 & 5.5846\\ \hline
80 & 0.05  & \bf{0.9956} & 4.7275 & 4.6576 & 5.4775\\ \hline
80 & 0.1   & \bf{0.9950} & 4.4705 & 4.3577 & 5.1277\\ \hline
80 & 0.5   & \bf{1.0041} & 2.9904 & 2.9675 & 3.9644\\ \hline
80 & 1     & \bf{1.0172} & 2.0566 & 2.0208 & 3.3470\\ \hline
\end{tabular}
\end{table}

In this section, we show the performance of various latent factor models based
on different hyper-parameter values to ensure that the WSVD model performs
consistently better than the other latent factor models. We modify two
important hyper-parameters ($k$ the number of latent factors and $\lambda$ the
regularization term) to several values and observe the corresponding RMSE
scores.  Specifically, we set the number of hidden factors to 10, 20, 40, and
80, and the regularization term to 0.001, 0.005, 0.01, 0.05, 0.1, 0.5, and 1.
We report the test RMSE scores on the MovieLens-100K dataset.

Table~\ref{tab:hyper-para-cmp} shows the test RMSE scores of the four latent
factor models under different hyper-parameter settings.  Here are our
observations.  First, in all our tried hyper-parameter settings, WSVD
outperforms the other models in terms of the test RMSE scores.  This shows the
effectiveness and generality of the WSVD model.  Second, the PMF model always
performs the worst.  This is probably because PMF model does not incorporate
the average rating and the biases of the users and the items.  Third, the RMSEs
becomes larger as the value of $k$ increases from 10, 20, 40, to 80.  This
seems to suggest that the number of latent factors to influence the ratings on
the MovieLens-100K dataset is a small number.  However, WSVD was influenced
minimally among the compared methods.  We will further discuss this observation
in Section~\ref{sec:learned-weights} Fourth, as the value of $\lambda$
increases from 0.001, 0.005, 0.01, 0.05, 0.1, 0.5, to 1, the test RMSEs of SVD
and SVD++ decreases, but the test RMSEs of the WSVD model increases.  This
implies that we need to apply a larger regularization term to the SVD and SVD++
models to prevent the overfitting problem.

%\subsubsection{Dive into the hyper-parameters of the WSVD model}
%
%\input{wsvd-para-cmp-table}
%
%In the above experiemnt, we found that, as the value of $\lambda$ increases,
%the test RMSEs of the WSVD model also increase.  This behavior is very
%different from the rest of the latent factor models.  We suspect that this is
%because the number of parameters in $\boldsymbol{P}$ and in
%$\boldsymbol{Q}$ is much larger than the number of parameters in $w$.  As a
%result, we should probably apply different regularization weights to them: the
%value of $\lambda_p$ and $\lambda_q$ should probably be larger than the value
%of $\lambda_w$.

\subsection{Learned weights} \label{sec:learned-weights}

\fig{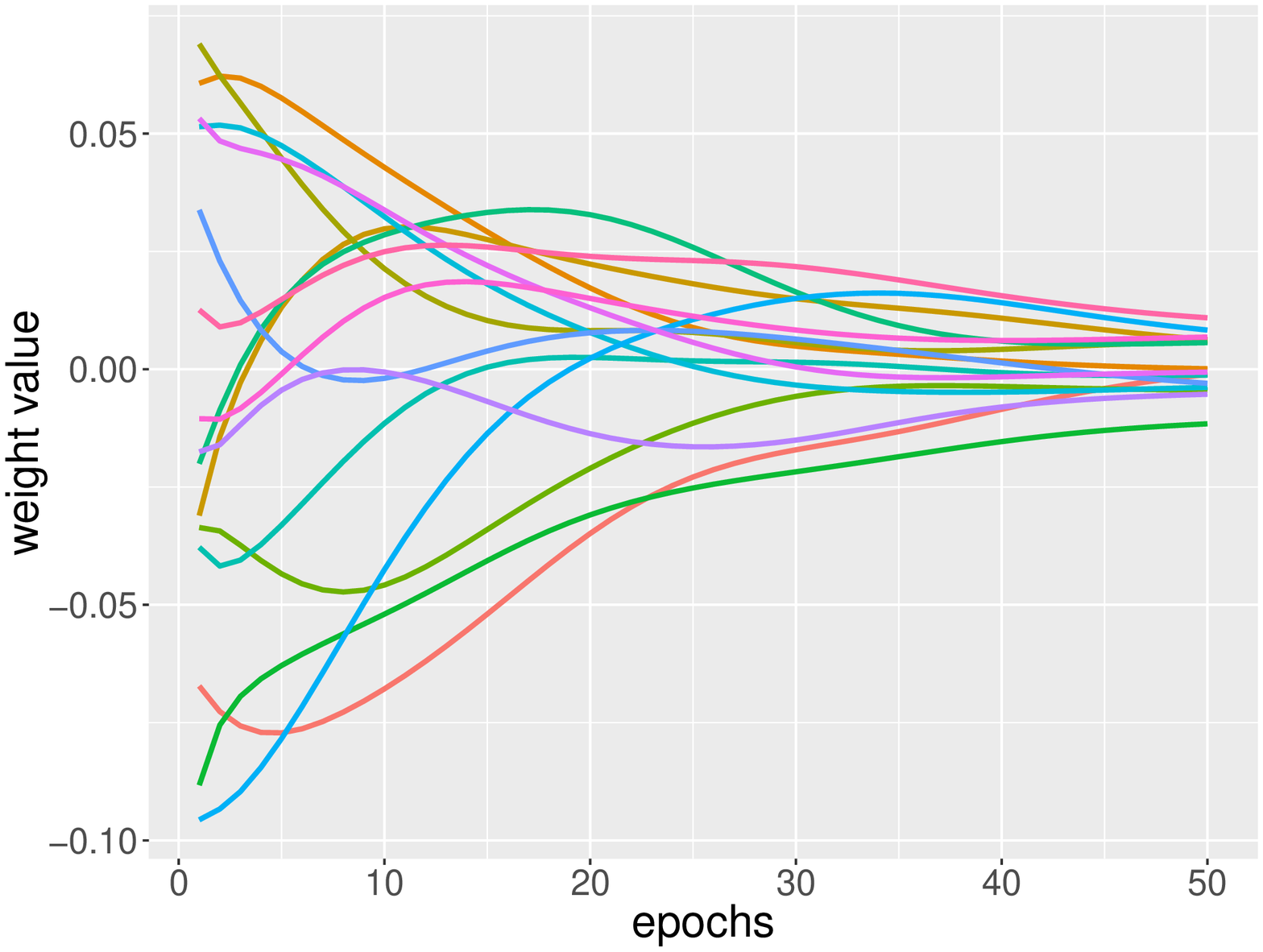}{Weights vs epochs (using the MovieLens-100K dataset)}{fig:weight-vs-epoch}{3.5}

\begin{table}[tb]
\centering
\caption{The relative importance of the learned weights}
\label{tab:relative-ratios}
\resizebox{440px}{!}{
\begin{tabular}{r||rrrrrrrrrrrrrrr}
\hline
$r_i$ & $r_1$ & $r_2$ & $r_3$ & $r_4$ & $r_5$ & $r_6$ & $r_7$ & $r_8$ & $r_9$ & $r_{10}$ & $r_{11}$ & $r_{12}$ & $r_{13}$ & $r_{14}$ & $r_{15}$ \\\hline\hline
val & -10.9 & 1 & 61.7 & 58.9 & -43.7 & -114.2 & 55.7 & 12.3 & -38.1 & 81.9 & -29.6 & -52.3 & -6.1 & 67.3 & 107.6 \\\hline
\end{tabular}
}
\end{table}

We report the learned weights of the latent factors in this section.  We used
the MovieLens-100K as the experimental dataset.

Figure~\ref{fig:weight-vs-epoch} exhibits the learning process of the 15 weights
on the 15 latent factors as the epoch goes from 1 to 50.  To show the relative
importance of each latent factor, we compute the relative importance $r_i$ of each
weight $w_i$ by Equation~\ref{eq:relative-strength}.

\begin{dmath} \label{eq:relative-strength}
r_i = w_i / \min_{\forall j}(|w_j|)
\end{dmath}

Table~\ref{tab:relative-ratios} lists the relative importance of the latent
factors.  As demonstrated, the important latent factors could be tens to
hundreds of times more influential than the less important latent factors.
This probably explains one observation in Section~\ref{sec:diff-hyper-para} --
the RMSE scores become much worse as the value of $k$ grows from 10, 20, 40, to
80 for the SVD, SVD++, and the PMF model, but the increment of the RMSE scores
of the WSVD model (as $k$ increases) is milder.  Since the WSVD model
automatically shrinks the influence of several latent factors, increasing the
number of latent factors may have smaller impact on the overall performance.

\subsection{Empirical training time}

\begin{table}[tb]
\centering
\caption{The average training time (in seconds) of one epoch of the compared methods}
\label{tab:exec-time}
\begin{tabular}{l||rrrr}
\hline
               & WSVD   & SVD    & SVD++    & PMF   \\\hline\hline
MovieLens-100K & 1.499  & 1.019  & 94.353   & 0.897 \\\hline
MovieLens-1M   & 15.957 & 10.975 & 2001.905 & 9.685 \\\hline
MovieLens-10M  & 157.736& 107.029& 21752.219& 88.417\\\hline
\end{tabular}
\end{table}
               
This section lists the empirical training time of the four compared methods.
We used the MovieLens-100K, MovieLens-1M, and the MovieLens-10M as the
benchmark datasets.  All the experiments reported in this subsection were
performed on a single computer with the Intel i7-6700 CPU 3.4GHz and 64GB RAM.
The OS is Ubuntu Linux 16.04.  We implemented all the methods in Python 3.6.2.

Table~\ref{tab:exec-time} lists the empirical training time.  As shown, the
training time of the WSVD, SVD, and the PMF models grows linearly with the size
of the dataset, and the training time of the SVD++ model grows super-linearly
with the size of the dataset.  The empirical training periods of all the
compared models follow our expectation, as analyzed in
Section~\ref{sec:model-analysis}.

\section{Related work} \label{sec:related-work}

Recommender systems usually aim to solve one of the two types of tasks: the
top-$k$ recommendation task and the user preference prediction task.  

For the top-$k$ recommendation task, the recommender system is expected to
suggest the $k$ items that best fit the target user's current needs based on
various clues, such as the property of the user, the property the items, users'
previous interactions with the items, and the context information. 

As for the preference prediction task, the system targets at predicting users'
preferences on \emph{all} items.  Thus, one may claim that the top-$k$
recommendation is a subset problem of the preference prediction task, since we
may always recommend the top-$k$ items based on the preference scores of all
items.  However, \cite{cremonesi2010performance} showed that utilizing the
preference prediction task to perform the top-$k$ recommendation may result in
sub-optimal recommendations, because minimizing the preference of \emph{all}
ratings may not necessarily improve the precision of the top-$k$ prediction.
Therefore, we may require different types of methods to address the two types
of tasks.

The preference prediction task becomes popular since the Netflix
prize~\cite{bennett2007netflix, bell2007lessons}, which is a challenge to
predict users' ratings on films.  This competition popularized the family of
the Matrix Factorization approaches, which generates the latent factors for
each user and each item and predict the preferences based on the inner product
of the latent factors.  The SVD model, SVD++ model, PMF model, and NNMF
(Non-Negative Matrix Factorization) model~\cite{mnih2008probabilistic,
koren2008factorization, koren2009matrix, lee2001algorithms} are the typical
representatives of this type of approach.  In this paper, we did not include
the NNMF model into the experiments because in practice its performance on the
preference prediction task is usually worse than the
others~\cite{kumar2009analysis}.

Since the preference score is a real number, some may argue that the preference
prediction task can be modeled as a regression task.  However, generating the
features based on purely on the known ratings $r_{ij}$ is not
straightforward~\cite{ma2008guide}.  Recently, \cite{rendle2010factorization}
proposed Factorization Machines (FM) that models the interaction of the
factorized parameters.  As a result, the FM model can process the dataset that
involves pairs of IDs (e.g., the user ID $i$ and the item ID $j$) that produce
sparse interactions (e.g., the rating $r_{ij}$).  Therefore, many Matrix
Factorization approaches are special cases of the Factorization Machines.
\cite{juan2016field} proposed the Field-aware Factorization Machines (FFM) to
further generalize the Factorization Machines such that the interactions
among more than two IDs can be included into the model.

\section{Discussion and future work} \label{sec:disc}

This paper proposed Weighted-SVD, a method to assign different weights to the
latent factors generated by the SVD model.  We showed the algorithms to learn
the user biases, item biases, the latent factor vectors for users and items,
and, perhaps more importatnly, the weights of the latent factors.  We found
that, compared to several MF approaches, such a method can better predict
users' preferences on items, based on the experiments on several open datasets.
In addition, based on the learned weights, the relative importance of the
significant latent factors could be tens to hundreds of times more influential
than the less significant factors, which suggests previous MF methods probably
over-simplified the scenario.  Since the weights on the less important latent
factors may shrink during the training, we may set the number of the latent
factors to a large number and let the model automatically discover the
appropriate dimensions of the latent factors.  This could be very beneficial,
since deciding an appropriate number of latent factor requires trial and error
by the other MF approaches.  We compared the model complexity and the training
time of several MF approaches.  The training time of the Weighted-SVD model is
only slightly larger than the SVD model, but the time complexity of training
for both the Weighted-SVD and the SVD models both grow linearly with the
training size.  

While we mainly discussed to extend the SVD model into weighted-SVD in this
paper, the weighting approach is very general.  Therefore, we may apply the
same technique to other MF methods, such as the SVD++ model and the
PMF model.

So far we assume that each hidden factor is independent from the other factors.
However, such an assumption may not always be correct.  We are interested to
investigate the approaches to include the dependency among the factors and the
weighting technique.  One possible direction is to integrate the Factorization
Machines with the weighting technique.

The current MF approaches model the interaction between a user's latent
factors and an item's latent factors by the inner product operation.  However,
it is possible that they are interacted in a different manner (e.g., with a
series of higher order operations).  Therefore, we are interested in designing
and experimenting different types of operations among the latent factors.
Deep learning and kernel methods are possible candidates.

Finally, we will open source a toolkit for the Weighted-SVD model so that
the research community of the recommender systems can be benefited.

\vskip 0.2in
\bibliographystyle{alpha}

\begin{thebibliography}{20}
\providecommand{\natexlab}[1]{#1}
\providecommand{\url}[1]{\texttt{#1}}
\expandafter\ifx\csname urlstyle\endcsname\relax
  \providecommand{\doi}[1]{doi: #1}\else
  \providecommand{\doi}{doi: \begingroup \urlstyle{rm}\Url}\fi

\bibitem[Bell and Koren(2007)]{bell2007lessons}
Robert~M Bell and Yehuda Koren.
\newblock Lessons from the netflix prize challenge.
\newblock \emph{ACM SIGKDD Explorations Newsletter}, 9\penalty0 (2):\penalty0
  75--79, 2007.

\bibitem[Bennett, Lanning(2007)]{bennett2007netflix}
James Bennett, Stan Lanning.
\newblock The netflix prize.
\newblock In \emph{Proceedings of KDD Cup and Workshop}, volume 2007, page~35.
  New York, NY, USA, 2007.

\bibitem[Chen et~al.(2011)]{chen2011collabseer}
Hung-Hsuan Chen, Liang Gou, Xiaolong Zhang, and C.~Lee Giles.
\newblock {CollabSeer: a search engine for collaboration discovery}.
\newblock In \emph{Proceedings of the 11th Annual International ACM/IEEE Joint
  Conference on Digital Libraries}, pages 231--240, Ottawa, Canada, 2011. ACM.

\bibitem[Chen et~al.(2015)]{chen2015expertseer}
Hung-Hsuan Chen, Alexander G.~Ororbia II, and C.~Lee Giles.
\newblock {ExpertSeer: a keyphrase based expert recommender for digital
  libraries}.
\newblock \emph{arXiv:1511.02058}, 2015.

\bibitem[Cremonesi et~al.(2010)]{cremonesi2010performance}
Paolo Cremonesi, Yehuda Koren, and Roberto Turrin.
\newblock Performance of recommender algorithms on top-n recommendation tasks.
\newblock In \emph{Proceedings of the Fourth ACM Conference on Recommender
  Systems}, pages 39--46. ACM, 2010.

\bibitem[Guo et~al.(2013)]{guo2013novel}
G.~Guo, J.~Zhang, and N.~Yorke-Smith.
\newblock A novel bayesian similarity measure for recommender systems.
\newblock In \emph{Proceedings of the 23rd International Joint Conference on
  Artificial Intelligence (IJCAI)}, pages 2619--2625, 2013.

\bibitem[Harper and Konstan(2016)]{harper2016movielens}
F.~Maxwell Harper and Joseph~A. Konstan.
\newblock The movielens datasets: History and context.
\newblock \emph{ACM Transactions on Interactive Intelligent Systems},
  5\penalty0 (4):\penalty0 19, 2016.

\bibitem[Huang et~al.(2014)]{huang2014refseer}
Wenyi Huang, Zhaohui Wu, Prasenjit Mitra, and C.~Lee Giles.
\newblock {RefSeer: a citation recommendation system}.
\newblock In \emph{Proceedings of the 14th ACM/IEEE-CS Joint Conference on
  Digital Libraries}, pages 371--374, London, UK, 2014. IEEE Press.

\bibitem[Juan et~al.(2016)]{juan2016field}
Yuchin Juan, Yong Zhuang, Wei-Sheng Chin, and Chih-Jen Lin.
\newblock Field-aware factorization machines for ctr prediction.
\newblock In \emph{Proceedings of the 10th ACM Conference on Recommender
  Systems}, pages 43--50. ACM, 2016.

\bibitem[Koren(2008)]{koren2008factorization}
Yehuda Koren.
\newblock Factorization meets the neighborhood: a multifaceted collaborative
  filtering model.
\newblock In \emph{Proceedings of the 14th ACM SIGKDD international conference
  on knowledge Discovery and Data Mining}, pages 426--434. ACM, 2008.

\bibitem[Koren et~al.(2009)]{koren2009matrix}
Yehuda Koren, Robert Bell, and Chris Volinsky.
\newblock Matrix factorization techniques for recommender systems.
\newblock \emph{IEEE Computer}, 42\penalty0 (8), 2009.

\bibitem[Kumar(2009)]{kumar2009analysis}
Aswani~Ch Kumar.
\newblock Analysis of unsupervised dimensionality reduction techniques.
\newblock \emph{Computer science and information systems}, 6\penalty0
  (2):\penalty0 217--227, 2009.

\bibitem[Lee and Seung(2001)]{lee2001algorithms}
Daniel~D Lee and H~Sebastian Seung.
\newblock Algorithms for non-negative matrix factorization.
\newblock In \emph{Advances in Neural Information Processing Systems}, pages
  556--562, 2001.

\bibitem[Liu et~al.(2010)]{liu2010personalized}
Jiahui Liu, Peter Dolan, and Elin~R{\o}nby Pedersen.
\newblock Personalized news recommendation based on click behavior.
\newblock In \emph{Proceedings of the 15th International Conference on
  Intelligent User Interfaces}, pages 31--40, Hong Kong, China, 2010. ACM.

\bibitem[Ma(2008)]{ma2008guide}
Chih-Chao Ma.
\newblock A guide to singular value decomposition for collaborative filtering,
  2008.

\bibitem[Massa and Avesani(2007)]{massa2007trust}
Paolo Massa and Paolo Avesani.
\newblock Trust-aware recommender systems.
\newblock In \emph{Proceedings of the 2007 ACM Conference on Recommender
  Systems}, pages 17--24. ACM, 2007.

\bibitem[Mnih and Salakhutdinov(2008)]{mnih2008probabilistic}
Andriy Mnih and Ruslan~R. Salakhutdinov.
\newblock Probabilistic matrix factorization.
\newblock In \emph{Advances in Neural Information Processing Systems}, pages
  1257--1264, 2008.

\bibitem[Rendle(2010)]{rendle2010factorization}
Steffen Rendle.
\newblock Factorization machines.
\newblock In \emph{The 10th IEEE International Conference on Data Mining},
  pages 995--1000. IEEE, 2010.

\bibitem[Ricci et~al.(2015)]{ricci2015recommender}
Francesco Ricci, Lior Rokach, Bracha Shapira, and Paul~B Kantor.
\newblock \emph{Recommender systems handbook}.
\newblock Springer, 2015.

\bibitem[Tang et~al.(2008)]{tang2008arnetminer}
Jie Tang, Jing Zhang, Limin Yao, Juanzi Li, Li~Zhang, and Zhong Su.
\newblock Arnetminer: extraction and mining of academic social networks.
\newblock In \emph{Proceedings of the 14th ACM SIGKDD International Conference
  on Knowledge Discovery and Data Mining}, pages 990--998, Las Vegas, USA,
  2008. ACM.

\end{thebibliography}

\end{document}